\documentclass[12pt]{article}

\usepackage{graphicx}
\usepackage[super]{natbib}
\usepackage{times}
\usepackage{amssymb}
\usepackage{setspace}

\def\mass#1{10^{#1}M_\odot}
\def\MLK{$M/L_K$}
\def\Msun{\hbox{$M_\odot$}}

\newcommand\aj{Astron. J.}
\newcommand\apj{Astrophys. J.}
\newcommand\apjl{\apj\ (Letters)}
\newcommand\mnras{Mon. Not. R. Astron. Soc.}
\newcommand\apjs{\apj\ Supp.}
\newcommand\nat{Nature}
\newcommand\aap{Astron. Astrophys.}

\newenvironment{natabstract}{
\begin{quote} \bf}
{\end{quote}}

\def\altaffilmark#1{$^{#1}$}

\bibpunct{}{}{,}{s}{}{,}

\author{Karl Glazebrook\altaffilmark{1},  Roberto G. Abraham\altaffilmark{2},
Patrick J.~McCarthy\altaffilmark{3}, \\
Sandra Savaglio\altaffilmark{1,4}, Hsiao-Wen Chen\altaffilmark{5,6}, 
David Crampton\altaffilmark{7}, \\
Rick Murowinski\altaffilmark{7}, 
Inger J{\o}rgensen\altaffilmark{8}, Kathy Roth\altaffilmark{8}, \\
Isobel Hook\altaffilmark{9}, 
Ronald O. Marzke\altaffilmark{10}, R. G. Carlberg\altaffilmark{2}
\\ \\
\normalsize{$^{1}$Department of Physics \& Astronomy, Johns Hopkins University,}\\
\normalsize{ 3400 North Charles Street, Baltimore, MD 21218-2686.}\\
\normalsize{$^{2}$Department of Astronomy \& Astrophysics,  University of Toronto,} \\
\normalsize{60 St. George Street,Toronto, ON, M5S~3H8, Canada}\\
\normalsize{$^{3}$Observatories of the Carnegie Institute of Washington, }\\
\normalsize{Santa Barbara Street, Pasadena, CA~9110}\\
\normalsize{$^{4}$On leave of absence from Osservatorio Astronomico di Roma, Italy}\\
\normalsize{$^{5}$Center for Space Research,  Massachusetts Institute of Technology,} \\
\normalsize{Cambridge, MA 02139-4307}\\
\normalsize{$^{6}$Hubble Fellow}\\
\normalsize{$^{7}$Herzberg Institute of Astrophysics, National Research Council, }\\
\normalsize{5071 West Saanich Road, Victoria, British Columbia, V9E~2E7, Canada.} \\
\normalsize{$^{8}$Gemini Observatory, Hilo, HI 96720}\\
\normalsize{$^{9}$Department of Astrophysics, Nuclear \& Astrophysics Laboratory,}\\
\normalsize{Oxford University, Keble Road, Oxford OX1 3RH, U.K.}\\
\normalsize{$^{10}$Dept. of Physics and Astronomy, San Francisco State University,}\\
\normalsize{1600 Holloway Avenue,  San Francisco, CA 94132 } 
\\
{\it Nature, in press}
}

\title{\bf The abundance of massive galaxies 3--6 billion years\\
after the Big Bang.}

\begin{document}

\maketitle

\begin{natabstract}
A fundamental prediction of our current hierarchical paradigm of
galaxy formation is that massive galaxies form from an assembly of
smaller units\citep{Blum84}. The most massive objects form last, driven by the merging
history of their dark matter haloes. The model succeeds in describing the clustering
of galaxies\citep{Sper03}, but the evolutionary history of massive
galaxies, as revealed by their visible stars and gas,
presents problems.  Near-infrared light allows us to measure the stellar
masses of high-redshift galaxies\citep{Brin00} and deep multi-colour images
indicate that a large fraction of the stellar masses in massive galaxies form 
in the first 5 Gyr\citep{Dick03,Font03,Drory01,Franx03}; but uncertainties remain due
the lack of spectra to confirm the redshift and the role of obscuration. Here we
report on the Gemini Deep Deep Survey, the deepest spectroscopic redshift survey ever
undertaken which probes the most massive and quiescent galaxies back to an era only 3 Gyr 
after the Big Bang. We find that at least two thirds of massive galaxies have appeared since this era, but also that a significant fraction is already in place indicating a suprisingly slow rate of decline as we look back in time.
\end{natabstract}

Star-formation rates  in the high redshift ($z>1$) Universe have been probed by
measuring rest-frame ultraviolet (UV) emission\citep{Madau96} of galaxies.
The UV is dominated
by newly formed massive stars but can be highly obscured\citep{Stei99}.
Moreover star-formation in
galaxies is a stochastic phenomenon and correlates poorly with mass:
the most massive galaxies in the local universe are
giant ellipticals and they have very weak UV emission \citep{Lotz00}
which makes them difficult to detect at $z>1$.
Direct determination of dynamical masses of galaxies requires spatially resolved
velocity measurements of galaxies --- this has been done out to
$z\sim 1$ \citep{Vogt97,Geb03} but such observations are very challenging
at higher redshifts due to signal-to-noise limitations.

An alternate approach is to infer the total mass in stars from light which traces 
accumulated stellar populations.
Studies\citep{Brin00} have shown that 
stellar mass correlates extremely well with dynamical
mass out to $z=1$. Stellar mass evolution can be predicted by
using galaxy formation models based on cosmological numerical simulations
augmented with analytical star-formation recipes,\citep{K99}
which  adopt simplified prescriptions for various heating and cooling processes
in the interstellar medium\citep{K99,Baugh03}. 

Deep near-infrared data provides a window on stellar masses of galaxies at
high redshift.
This light is dominated by the
old, evolved stellar populations in galaxies and is little affected by
transient star-formation \citep{RR93} or dust obscuration.  To a good approximation
the total $K$-band light traces the accumulation of stellar mass;
equivalently one can say that the stellar mass-to-light ratio ($M/L_K$)
is nearly constant with little dependence on the previous
star-formation history (SFH). In fact rest-frame  \MLK\ varies only by a factor of two
between extremely young and extremely old galaxy stellar populations, in contrast 
$M/L_B$ can vary by more than a factor of ten \citep{Bell03} and
$M/L_{UV}$ varies tremendously (reaching as low as 1\% of
the solar value) because the UV light is dominated by the
instantaneous star-formation rate (SFR) rather than the stellar
mass. Even at redshifts up to $z=2$
the {\it observed}-frame $K$-band probes the rest-frame $R$-band which is
still dominated by old stellar populations.  The $M/L_R$ variation is still less than a factor of four,
partly due to the youth of the Universe at that redshift.

We use the Gemini Deep Deep Survey (GDDS)\citep{Abr04} to define a new sample  
of 150 galaxies with $K<20.6$ and $0.8<z<2$ located in four independent 30 arcmin$^2$
fields. The spectroscopic identification completeness is 89\%. The determination
of stellar masses for each galaxy 
follows standard multi-color stellar population fitting techniques; this approach
is fairly general and robust and is detailed in the Methods appendix. We find that the $K$-band light
traces the stellar mass quite well; evolutionary changes in galaxy numbers with redshift are much
more important than changes in color. The findings
presented below are simply driven by the presence of numerous $K\sim
20$ galaxies at $z>1.5$ which must be massive objects.

Figures 1 \& 2  illustrate the nature of the most massive galaxies. They are found by the GDDS 
to $z=2$ at $K\sim 20$, even when they
are very red ($I-K>4$).  The red galaxies are predominantly red due to old stellar populations
(these are GDDS spectral classes\citep{Abr04} `001' and `011'  showing  photospheric features from evolved stars) 
and not dust reddening of young lower mass galaxies. The red,
old galaxies make a large contribution (30\%) to the stellar mass in the Universe in the redshift range
$1.2<z<1.8$.  Also at $z>1$ there are a number of blue galaxies despite this being
a $K$-selected sample. These correspond to {\em massive} star-forming
galaxies which have high metal abundances\citep{Savaglio04}. 

The cumulative stellar mass density per unit volume  in each
redshift bin down to various mass thresholds (Figure~3; Table~1) is computed following the standard
$V/V_{max}$ formalism (this corrects for the 
smaller redshift ranges covered by fainter objects\citep{Vmax}) for a $K<20.6$ limit
and weighting by the sampling. We use
K-corrections from our individual spectral energy distribution (SED) fits, but the results
do not depend strongly on the details of the K-correction.
Although a $K$-selected sample is a good
proxy for a mass-selected sample, it is still necessary to consider
incompleteness as a function of $M/L$. To do this we compute the
maximum possible $M/L$ at each redshift for a model galaxy as old as
the Universe which formed all its stars at once (a `Simple Stellar Population' or SSP). This $M/L$ is
converted to a mass limit (via the $K$-band flux limit and the
K-correction) and is shown in Figure~1. We are complete above this mass
limit; bluer objects can be seen below this limit. Bins which might miss high $M/L$ objects
are plotted as {\em lower limits} in Figure 3. The error bars are calculated from shot noise on the
number of galaxies in each bin. Of course these do not include the effects of large-scale
structure, but we believe
these are not significant because our GDDS fields are large, were selected from even larger area images
in regions near average density\citep{Abr04} (i.e. neither highly over-dense nor under-dense) and because
our colour-dependent weights normalise to the full imaging
area (554.7 arcmin$^2$) which would counteract the additional clustering of
red objects. Finally we have performed the check of splitting the sample
by different fields (different independent sight-lines). The same general results are found for these
albeit with larger errors. We have also assessed the effect of the
mass fitting errors on the mass densities with our
Monte-Carlo methods; this is not a significant
source of error. In every bin galaxies with spectroscopic redshifts
dominate the mass budget except for the $1.6<z<2$ and $M>\mass{10.8}$
bin,  where the spectroscopic completeness is only 50\% and we augment our spectroscopy with
photometric redshifts. Analysing the sub-sample with only spectroscopic redshifts results in no significant changes to any bin except for this one (which is thus 0.3 dex lower); the essential scientific result is unchanged.

There is a  clear decline in stellar mass locked up in the most massive galaxies. This trend is in accord with 
ideas of gradual assembly. However it declines only slowly towards  high redshift and, surprisingly,
the massive galaxies do not
decline more rapidly than the whole
population. We note that galaxies with $M>\mass{10.8}$, 
are brighter than our flux limits (for any possible $M/L$ value) throughout our redshift range; 
a regression on these gives an acceptable fit for $\rho\propto (1+z)^{-1.7 \pm 1.6}$. 
At $z=1$ the mass densities for 
the $M>\mass{10.8}$ sample are
38 $(\pm 18)$\% of their  local value\cite{Cole01}; at $z=1.8$ this becomes 16 $(\pm 6)$\%. These results
for the most massive galaxies are 
consistent with previous Hubble Deep Field South photometric redshift determinations\cite{Dick03,Font03},
but inconsistent with the more rapid  decline (factor of $\sim 6$  over $0<z<1$)
found with photometric redshifts by the Munich Near-Infrared Cluster Survey\citep{Drory01}
(however we note that their SED fitting forces maximally old galaxies at all redshifts, which could be
problematic).  Overlaid on Figure 3 (shaded region) is the range of estimates for the
growth of total stellar mass based on the integral of the observed rest-UV derived
SFR-$z$ relationship. (These are based on the an analytic fit\cite{Cole01} of
points from Figure 9 of \citet{Stei99},
both with and without  extinction correction, integrated using PEGASE.2\citep{PEGASE}.) 
Our uppermost points represent lower limits on the 
total stellar mass density and we confirm previous findings that an extinction
correction is essential for UV SFR estimates to be consistent
with stellar mass measurements\cite{Dick03,Cole01}.

Theoretical models of galaxy formation convert gas into stars in dark matter halos using semi-empirical recipes and must satisfy three key observables: the first is the distribution 
of local galaxy luminosities, the second is that of galaxy colors and the third is the abundance of massive galaxies at high redshift. The last of these has until now been the most difficult to measure from observations. In Figure 3 we plot the abundance predictions of the `GALFORM' models\citep{Baugh03,Gran00} which satisfy the first two constraints; it is evident in this model
that massive galaxies disappear much more rapidly than we see in our data; this is
because  the model stellar mass build-up traces
the merging of cold dark matter haloes. In particular GALFORM has a strong dependence
of evolutionary rate on mass which is not seen in our data. A similar theoretical comparison was made 
on the Hubble Deep South photometric data with similar conclusions\citep{Font03}.

We note that our measured abundance of massive galaxies at $z\sim 2$ does not violate the
{\em boundary constraints} of cold dark matter models, by which we mean taking the predicted
abundance of $z=2$ massive dark matter halos \citep{ST99} and scaling by the cosmological
baryon $/$ dark matter density \cite{Sper03} ($\Omega_b/\Omega_m=0.17$). 
This gives a predicted baryonic mass
density of $\simeq 10^8$  $\Msun$ Mpc$^{-3}$ in haloes with baryonic masses $>\mass{11}$, a 
factor of ten above our measurements. Thus the models can match our 
densities if only $\sim$10\% of the baryons in massive haloes are converted in 
to stars by $z=2$. We note though that in the Universe today, the stellar density\citep{Cole01} 
$\Omega_\star = 0.004$ gives $\Omega_\star/\Omega_b$ = 0.1; this would imply 
massive haloes at $z=2$ have managed to convert baryons into stars with the same overall efficiency as the {\it average\/} Universe had achieved by $z=0$. Massive haloes must have much greater star-formation efficiencies at earlier times.

Models have been proposed \citep{K99,Somer01} which adjust  the star-formation histories in this way and which would be more in accord with our findings of a high abundance of massive galaxies at early times. However they  fail to match existing data on galaxy colours\citep{Somer04,Stan04} or the galaxy luminosity function\citep{K99} so we defer detailed comparison to future work.

What is {\em observationally} clear is that we have measured directly the abundance of 
massive galaxies out to $z=2$ for the first time from a deep and relatively wide-area survey with secure 
redshifts and spectroscopic classifications for galaxies. We find (with 99\% confidence) that at least two-thirds of the mass in these objects has formed since $z=1.8$. However
we do not find a more rapid evolution of the giant population compared to the
Universal average as would be expected if stellar mass grew simply proportional to dark matter assembly.
At $z\lesssim 2$ we find much of the mass is in galaxies with old stellar populations;
these objects are 
plausible precursors of modern massive elliptical galaxies.  This new data is broadly consistent with 
developing ideas that models of galaxy formation have to be tuned towards 
earlier star-formation in high-mass haloes. The abundance of massive galaxies at $z\lesssim 2$
is now firmly established and will  provide the
vital missing leg in the tripod of key observations required to understand how galaxies form.

\section*{Methods}

\subsection*{Stellar Mass Fitting}

We adopt
a cosmology\citep{Sper03} of $\Omega_{m}=0.3$,  $\Omega_{\Lambda}=0.7$, $H_{0}=70$
${\rm km\,s}^{-1}\,{\rm Mpc}^{-1}$ and use Vega magnitudes. 
For the mass-function analysis we use GDDS galaxies with $K<20.6$ and $0.8<z<2$ (150 galaxies; 89\% spectroscopic completeness based on \citet{Abr04} identification confidence classes $>=2$; galaxies without
spectroscopic redshifts are assigned to their photometric redshifts). To account for the
higher priority given to red galaxies in the slit-mask design we use the `sampling weights' from
\citet{Abr04}  which are the selection probability for spectroscopy as a function of $I-K$ and $K$. 

For each galaxy we derive the most likely stellar mass and a range of uncertainty.
These are derived by evaluating against  model optical-IR
spectral-energy distributions (SED) 
to determine the $M/L_K$ ratio, and hence the mass. We note that a
variety of approaches to accomplish this have been described in the literature
which vary in the level of detail in which they treat star
formation. For example, \citet{Cole01} used a simple set of
monotonic star-formation histories (SFH) with a varying e-folding time-scale
and a fixed dust law. A potential problem is posed by the fact that
real galaxies have more complex SFHs, for example a recent starburst
can make an old galaxy temporarily bluer and lead to an underestimate
of $M/L$ using this method. One approach to account for this effect is
to introduce a second young SED component\cite{Font03}
superimposed on the old population; this can be computationally
expensive depending on the amount of freedom allowed for the second
component.

In a spectroscopic sample there is no degeneracy between SED model
fitting and photometric redshift (which is also based on SED
fitting). We adopt two-component modelling (using PEGASE.2
\citep{PEGASE} to calculate spectra) in order to be able to assess the
possible biases due to starbursts on the calculated masses and allow a
range of dust extinction ($0\le A_V \le 2$ mag) and metallicity
($0.0004\le Z\le0.02$). The primary component is modelled using a star-formation rate 
$\hbox{SFR} \propto\exp(-t/\tau)$ with $\tau=$0.1, 0.2, 0.5, 1, 2, 4,
8 \& 500 Gyr (the first approximates an instantaneous starburst and
the last a constant SFR).  The secondary component is a starburst,
modelled with a $\tau=0.1$ Gyr exponential, which can occur at any time
and have a mass between $10^{-4}$ and twice that of the primary component. 
The model
age is constrained to be less than that of the Universe at the appropriate redshift,
but any formation epoch is allowed.
Like all studies of high redshift star-formation, we must assume a Universal
Initial Mass Function (IMF). We primarily
use the BG03 IMF\citep{BG03} which fits local cosmic luminosity
densities well; it has a similar high-mass slope as the classical Salpeter 
IMF\cite{Salp55} but with
a more realistic break at 0.5\Msun. Stars with masses below the break never contribute
significantly to optical/IR light so the overall effect is to re-scale the total stellar masses
to more reasonable values. Similarly we  re-scale literature Salpeter based 
numbers\citep{Dick03,Font03,Cole01}
to the BG03 IMF  
using $M_{SP}=1.82 M_{BG}$. Our results are robust for any reasonable choice
of IMF with a similar slope; to illustrate this we have also calculated masses using
the popular Kennicutt  IMF\cite{Kenn83}.

Our approach is to find via exhaustive grid search all models consistent with
the $VIz'K$ photometry in the observed frame
of each galaxy. This colour set is available for all galaxies and
covers rest-frame ultraviolet through near-infrared. 
We do not include the actual spectra (apart
from the redshift information) in the fits because of variable quality
and signal-to-noise, however we find the spectral classes are broadly consistent
with the best photometric SED fits. The full distribution function of allowed
masses were calculated by Monte-Carlo re-sampling the photometric
errors. The final masses and error bars represent the mean and standard
deviation of this full distribution function. Typically, we
find the masses are fitted to $\pm 0.17$ dex in the $K<20.6$ sample.  
The stellar masses are very robust against 
the details of the fitting.  Using
the Monte-Carlo machinery to investigate the effect of different assumptions about
metallicity, dust and bursts we find that the largest effect is due to
bursts. If we disallowed bursts then the masses typically decrease by
only 0.2 dex. Finally we note the variation in $M/L_K$ over the range 
$1<z<2$ is constrained by the age of the Universe (6--3 Gyr), typically the maximum range in
our sample is a factor of three.

\newpage

\medskip\noindent
\begin{small}
{\bf Correspondence} and requests for materials should be addressed to K.G. (kgb@pha.jhu.edu)
\\\\
{\bf Acknowledgments} Based on observations obtained at the Gemini
Observatory, which is operated by AURA under a co-operative agreement
with the NSF on behalf of the Gemini partnership: NSF (U.S.), PPARC
(U.K.), NRC (Canada), CONICYT (Chile), ARC (Australia), CNPq (Brazil)
and CONICET (Argentina). Also based on
observations made at the Las Campanas Observatory of the Carnegie Institution of Washington.
Karl Glazebrook \& Sandra Savaglio
acknowledge generous funding from the David and Lucille Packard
Foundation. Roberto Abraham gratefully acknowledges receipt of funding from NSERC
and from the Government of Ontario through
a Premier's Research Excellence Award. Hsiao-Wen Chen acknowledges support by NASA
through a Hubble Fellowship grant from the Space Telescope
Science Institute, which is operated by the Association of Universities 
for Research in Astronomy, Incorporated, under NASA contract.
\\\\
{\bf Competing interests statement}\quad The authors declare that they have no competing financial interests.
\\\\
\end{small}

\begin{table}[htdp]
\begin{center}
\begin{tabular}{ccccccccc}
\hline\hline
 & & \multicolumn{3}{c}{BG03 IMF} & \multicolumn{3}{c}{Kennicutt IMF} & \\
\hline
$\log M_{lim}$ & $z$ range & $\log\rho$ &$\log \rho_{lo}$ & $\log\rho_{hi}$ 
& \quad\quad  $\log\rho$ &$\log \rho_{lo}$ & $\log\rho_{hi}$& Complete? \\
\hline
10.2 &   0.8--1.1 &  7.92 &  7.79  &  8.02 & \quad\quad    7.82 &  7.67  &  7.92 &   Y \\ 
10.2 &   1.1--1.3 &  7.56 &  7.43  &  7.66 & \quad\quad    7.47 &  7.34  &  7.58 &   N \\ 
10.2 &   1.3--1.6 &  7.82 &  7.73  &  7.90 &  \quad\quad   7.71 &  7.62  &  7.79 &   N \\ 
10.2 &   1.6--2.0 &  7.43 &  7.30  &  7.54 &  \quad\quad   7.34 &  7.21  &  7.44 &   N \\ 
\hline
10.5 &   0.8--1.1 &  7.86 &  7.70  &  7.97 & \quad\quad    7.72 &  7.54  &  7.84 &   Y \\ 
10.5 &   1.1--1.3 &  7.42 &  7.25  &  7.55 &  \quad\quad   7.34 &  7.17  &  7.47 &   Y \\ 
10.5 &   1.3--1.6 &  7.73 &  7.63  &  7.82 &  \quad\quad   7.62 &  7.51  &  7.71 &   Y \\ 
10.5 &   1.6--2.0 &  7.36 &  7.21  &  7.47 & \quad\quad    7.26 &  7.11  &  7.37 &   N \\ 
\hline
10.8 &   0.8--1.1 &  7.61 &  7.34  &  7.77 & \quad\quad   7.46 &  7.13  &  7.64 &   Y \\ 
10.8 &   1.1--1.3 &  7.21 &  6.94  &  7.37 & \quad\quad   6.92 &  6.49  &  7.14 &   Y \\ 
10.8 &   1.3--1.6 &  7.41 &  7.25  &  7.53 & \quad\quad    7.24 &  7.04  &  7.38 &   Y \\ 
10.8 &  1.6--2.0 &  7.24 &  7.06  &  7.37 & \quad\quad   6.96 &  6.74  &  7.11 &   Y \\ 
\end{tabular}
\caption{ Stellar Mass Density measurements in the universe as a function of galaxy mass threshold and redshift. $\rho$ is the stellar mass density in galaxies with stellar masses $>M_{lim}$ in $\Msun$ Mpc$^{-3}$. The values of $\rho_{hi}$ and $\rho_{lo}$ are the 1$\sigma$ upper and lower limits from
counting statistics. BG03 IMF mass thresholds and mass
densities can be multiplied by 1.82 to convert to the Salpeter IMF.}
\end{center}

\end{table}

\clearpage\pagestyle{empty}

\begin{figure}
\centerline{
\includegraphics[width=5.4in]{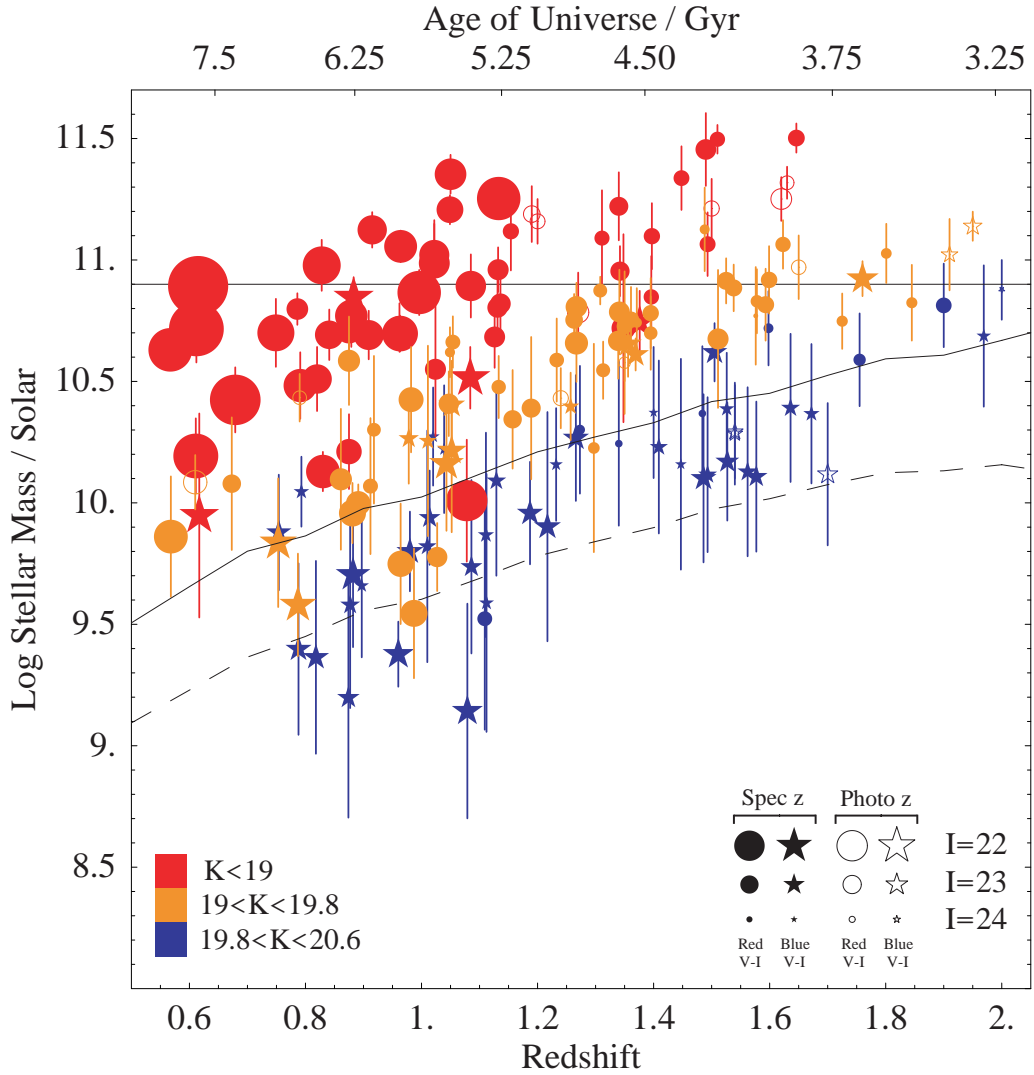}
}
\caption{\sf Stellar mass-redshift 
distribution for our galaxies. The 1$\sigma$ error bars come from  the Monte-Carlo mass fitting.
Symbol colours code the observed  $K$-magnitude (see panel). Solid/open symbol shapes denote 
spectroscopic/photometric redshifts respectively. Circle/star symbols denote objects redder/bluer in $V-I$ than a model Sbc galaxy template\cite{CWW} respectively.
Symbol size is keyed to the $I$-band
magnitude. The horizontal line denotes the characteristic Schechter mass scale 
in the local Universe\citep{Cole01}.
The solid curve shows how the $K$-flux limit translates
into a mass completeness limit for a maximally old Simple Stellar Population. The dashed curve shows an example mass limit for bluer objects (SFR$=$const. model).}
\end{figure}

\begin{figure}
\centerline{
\includegraphics[width=5.5in]{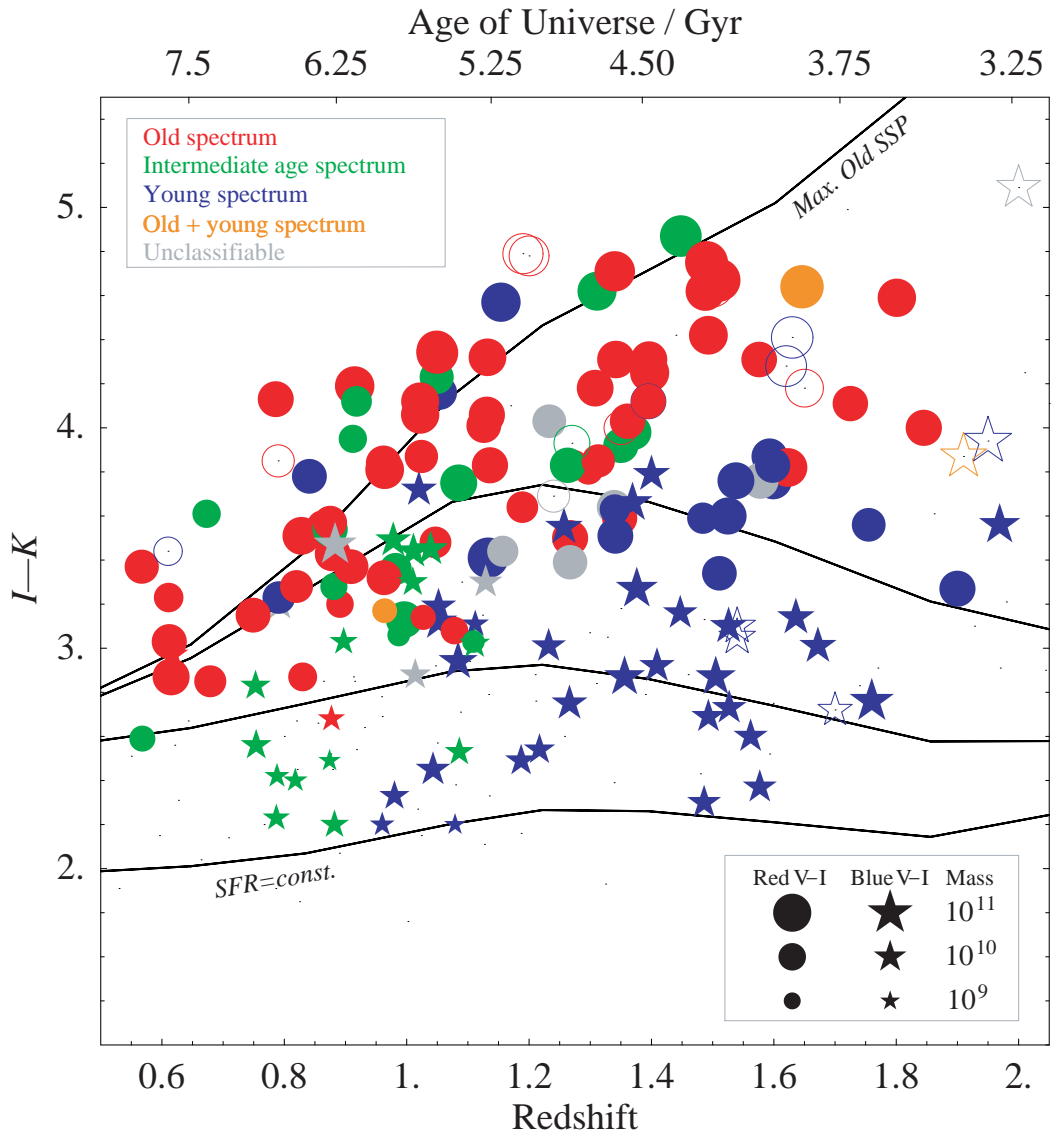}
}
\caption{ \sf Colour-redshift distribution for our galaxies. Observed frame $I-K$ colour is plotted with
symbol size keyed to mass, shape to
optical $V-I$ colour (as in Figure 1)
and  symbol colour keyed to the {\em spectra\/l} classification (see panel). Solid/open
symbols denote spectroscopic/photometric redshifts. Model tracks (solid lines) are shown ranging from a
a maximally old Simple Stellar Population (SSP) through to SFR$=$const., for synthetic solar metallictiy galaxies which form at $z=10$. (Note the points red-ward of the old track at $z=0.6$ may be artificially reddened by 
observational aperture/metallicity effects, this is insignificant for $z>0.8$.)
}
\end{figure}

\begin{figure}
\centerline{
\includegraphics[width=6.5in]{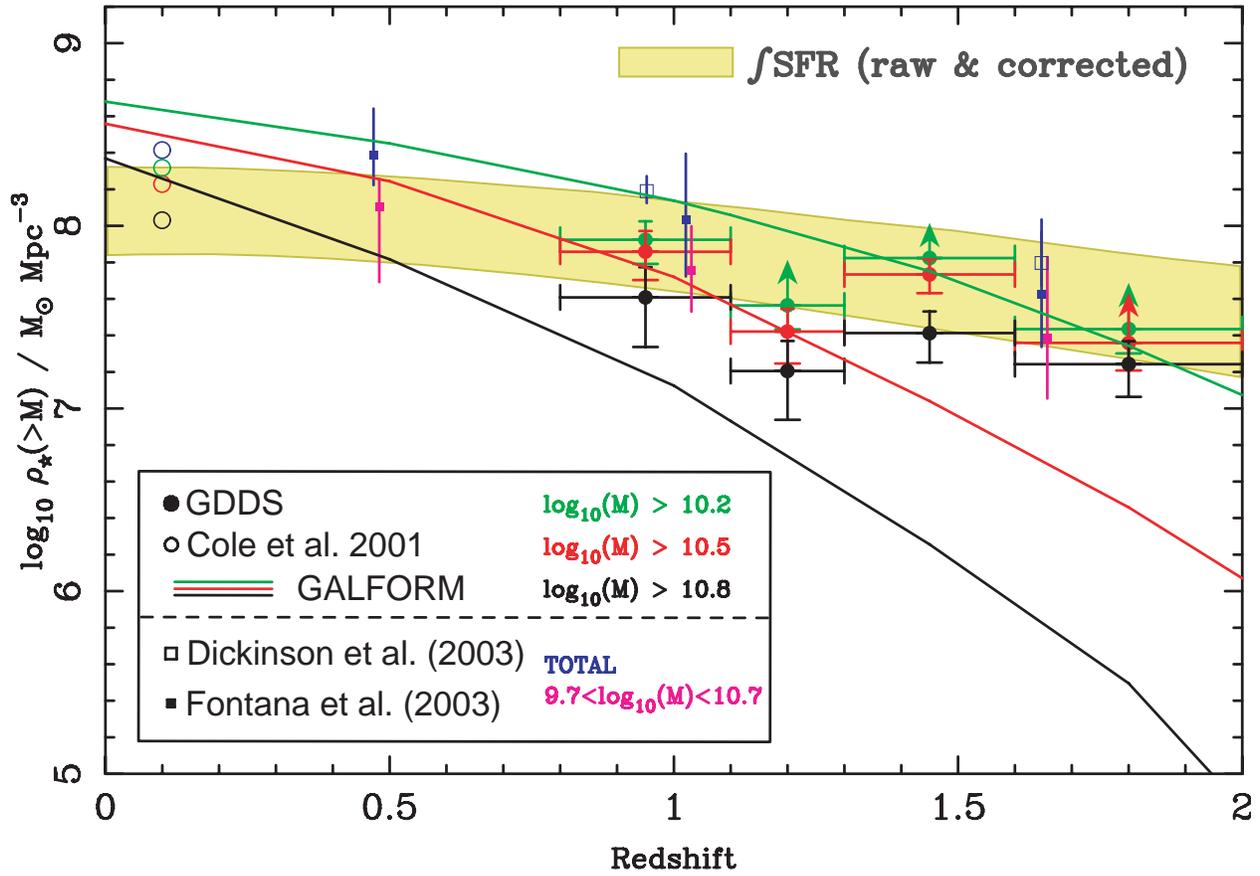}
}
\caption{ \sf Mass density in stars vs redshift. Values from our spectroscopic sample 
are compared with previous estimates which 
are all based on photometric redshifts\citep{Dick03,Font03} (except for the local $z=0.1$ point\citep{Cole01}). We plot the cumulative mass density of galaxies (converted to our Initial Mass Function choice; error bars are 1$\sigma$)
more massive than a  given mass threshold (see legend for threshold colour coding). Theoretical `GALFORM' models are also plotted (see text). The shaded region shows the
result of integrating the Universal ultraviolet-derived star-formation history with and without a dust 
extinction correction.
}
\end{figure}

\end{document}